\documentclass[fleqn,twoside]{article}
\usepackage{espcrc2}

\usepackage{graphicx}
\usepackage[figuresright]{rotating}

\newcommand{\AmS}{{\protect\the\textfont2
  A\kern-.1667em\lower.5ex\hbox{M}\kern-.125emS}}

\newcommand{\munuebar}{\rm{\mu_{\nu} ( \nuebar ) }}
\newcommand{\mub}{\rm{\mu_B}}
\newcommand{\enu}{\rm{E_{\nu}}}
\newcommand{\nuebar}{\bar{\nu_e}}
\newcommand{\nue}{\nu_e}
\newcommand{\munu}{\mu_{\nu}}

\title{
{\normalsize \hfill AS-TEXONO/04-03 \\
\hspace*{1cm} \hfill \today\\}
Neutrino Magnetic Moments: Status and Prospects
}

\author{Henry T. Wong\address[AS]{Institute of Physics,
        Academia Sinica, Taipei 11529, Taiwan.}}%
       
\begin{document}

\begin{abstract}
Finite neutrino magnetic moments are consequences
of non-zero neutrino masses. The particle physics foundations
of the subject are summarized. 
The astrophysical bounds  as well as
the results from recent direct experiments are
reviewed. Future projects and prospects 
are surveyed.
\vspace{1pc}
\end{abstract}

\maketitle

\section{INTRODUCTION}

The strong evidence of neutrino oscillations 
from the solar, atmospheric and long baseline
accelerator and reactor neutrino measurements
implies finite neutrino masses and mixings\cite{pdg04,nu04}.
Their physical origin and experimental consequences
are not fully understood.
Experimental studies on the neutrino properties
and interactions can shed light to these
fundamental questions and provide constraints to
the interpretations in the
future precision oscillation experiments.
New and improved neutrino sources and detector technologies
have to be developed in parallel for  such studies.

The couplings of neutrinos with the photons are 
generic consequences of finite
neutrino masses, and are one of the 
important intrinsic neutrino properties\cite{nuprop}
to explore.
The neutrino electromagnetic vertex can be
parametrized by terms with $\gamma_{\eta}$
and $\sigma_{\eta \xi}$ corresponding
to interactions without and with 
its spin, respectively identified as
the ``neutrino charge radius'' and 
``neutrino magnetic moments'',
the latter of which is the
subject of this review
\footnote{Proceedings to the XXIst International Conference on
Neutrino Physics and Astrophysics, Paris, 2004}.

\section{PARTICLE PHYSICS OVERVIEW}

The most general form for the effective
Lagrangian describing the spin
component of the neutrino electromagnetic vertex 
can be expressed as 
\begin{equation}
L = \frac{1}{2} \bar{\nu}_j 
\sigma_{\eta \xi} ( \beta_{ij} + \epsilon_{ij} \gamma_5 )
\nu_i  F^{\eta \xi} + h.c. 
\end{equation}
where $\epsilon_{ij}$ and $\beta_{ij}$ are respectively
the electric and magnetic dipole moments which couple
together the neutrino mass eigenstates $(\nu_i)_L$ and $(\nu_j)_R$,
resulting in a change of the spin-state.
Cases where $\nu_i = \nu_j$ and $\nu_i \neq \nu_j$ 
correspond to
{\it diagonal} and {\it transitional} moments, 
respectively.
Symmetry principles as well as neutrino properties
place constraints to the matrices
$\epsilon_{ij}$ and $\beta_{ij}$\cite{kaysernieves}.
For example, Majorana neutrinos require
$\epsilon_{ii}=\beta_{ii}=0$ which implies the diagonal
moments vanishes. The study of neutrino electromagnetic
properties is, therefore, 
in principle a way to distinguish
between Dirac and Majorana neutrinos.

The experimental observable ``neutrino magnetic moment'' 
($\munu$), usually expressed in units
of the Bohr magneton ($\mub$),  for
neutrinos with energy $E_{\nu}$ produced as 
$\nu_l$ at the source and after traversing a distance L
can be described by 
\begin{equation}
\munu^2 ( \nu_l , L , E_{\nu} ) =
\sum_{j} | \sum_{i} U_{li} ~ e^{-i E_{\nu} L } ~ \mu_{ij} | ^2  ~~ ,
\end{equation}
where $\mu_{ij} \equiv | \beta_{ij} - \epsilon_{ij} |$
and $U_{li}$ is the neutrino mixing matrix.
The observable
$\munu$ is therefore 
an effective and convoluted parameter
and the interpretations of
experimental results depend on 
the exact $\nu_l$ compositions at
the detectors. 
Accordingly, the $\munu$ limits
from reactor experiments are not identical
to those from $^8$B solar neutrino
experiments, which in turn are different
to those from $^7$Be solar neutrino experiments.

Given a specific model, $\munu$ can be calculated
from first principles.
Minimally-Extended Standard Model with massive 
Dirac neutrinos\cite{pdg04}  gives 
$\munu \sim 10^{-19} [ m_{\nu} / 1 \rm{eV} ] $
which is far too small to have any observable
consequences. Incorporation
of additional physics, such as 
Majorana neutrino transition moments or
right-handed weak currents, can significantly
enhance $\munu$ to the experimentally relevant
ranges\cite{pdg04,vogelengel}. Supersymmetry can also contribute
to the process, and the consequences from
models based on extra-dimensions were recently
discussed\cite{exdim}.

\begin{figure}[h]
\center
\includegraphics[width=6.5cm]{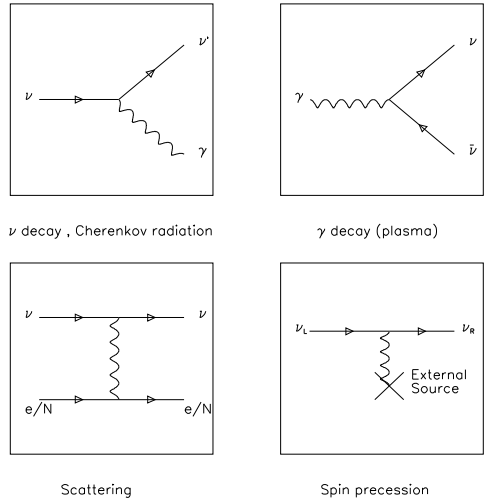}
\caption{
Schematic diagrams for the various
experimental manifestations of neutrino-photon
couplings.
}
\label{processes}
\end{figure}

Information on $\munu$ can be
derived from astrophysics arguments as
well as from direct laboratory experiments.
The various manifestations of the neutrino-photon
couples are shown Figure~\ref{processes}.
In particular, studies of neutrino-electron
scatterings are the most robust and established
methods.
A finite $\munu$ gives rise to an additional
contribution in the $\nu$-e scattering 
differential cross-section\cite{vogelengel}
\begin{equation} 
\label{eq::mm} 
( \frac{ d \sigma }{ dT } ) _{\munu}  ~ = ~
\frac{ \pi \alpha _{em} ^2  }{ m_e^2 }
 [ \frac{ 1 - T/E_{\nu} }{T} ] ~ \munu^2 
\end{equation}
where {\it T} the electron recoil energy, the
experimental measurable. 
The neutrino
radiative decay rate $\Gamma_{ij}$
for the process $\nu_i \rightarrow \nu_j + \gamma$
is related to $\mu_{ij}$ via\cite{ndecay}
\begin{equation} 
\Gamma_{ij} =
\frac{1}{8 \pi} \frac{( m_i^2 - m_j^2 ) ^ 3}{m_i^3}
\mu_{ij}^2  ~ ~ ,
\end{equation}
where $m_{i,j}$ are the masses for neutrino mass-eigenstates
$\nu_{i,j}$.

\section{ASTROPHYSICS BOUNDS}

Astrophysics bounds on $\munu$ were mostly
derived from the consequences from
a change of the neutrino spin-states 
in the astrophysical medium\cite{pdg04,raffeltbook}. 
These include studies
in the available
degrees of freedom in Big Bang Nucleosynthesis,
stellar cooling via plasmon decay, and the
cooling of supernova 1987a.
The typical range is 
$\munu (astro) < 10^{-10} - 10^{-12} \mub$.

The bounds, however, depends on modeling
of the astrophysical systems, as well as
on placing certain assumptions on the
neutrino properties. The supernovae cooling
arguments only apply for Dirac neutrinos
where the right-handed state is sterile
and can leave the astrophysical objects
readily. Another generic assumption is
the absence of other non-standard
neutrino interactions except for
an anomalous magnetic moment.
For more realistic studies, a global
treatment would be desirable, incorporating
oscillation effects, matter effects as well
as the complications due to interference
and competitions among various channels.

As a historical footnote, the spin-flavor
precession (SFP) mechanism, with or without matter
resonance effects in the solar medium,
has been used to explain solar neutrino
deficit\cite{sfp}. The solar $\nue$ would interact
with solar magnetic field $\rm{B_{\odot}}$ 
via its magnetic moment to
become $\nu_x (x$$\ne$$e)$. This scenario
is in fact compatible with all solar neutrino data.
The terrestrial KamLAND experiment, however,
recently confirmed the Large Mixing
Angle (LMA) parameter space
of the matter oscillation
scenario as {\it the} solution
for the solar neutrino problem\cite{pdg04,nu04}, such
that SFP can be excluded as
the dominant contribution in solar neutrino physics.
Conversely, coupling the LMA allowed region with the
recent KamLAND solar-$\nuebar$ bounds of
$\nuebar / \nu_{\odot} < 2.8 \times 10^{-4}$\cite{KLnuebar},
a constraint on $\rm{ \int \munu [ B_{\odot \perp } ] ~ dr }$
can be derived,
where $\rm{ B_{\odot \perp }}$  denotes
transverse component of the solar magnetic field. 
Recent work on the
modeling of $\rm{B_{\odot}}$\cite{bsolar}
turned this into bounds on  the magnetic
moments, also in the range of
$\munu ( ^8$B-$ \nu_{\odot} ) < 10^{-10} - 10^{-12} \mub$.

\section{RECENT RESULTS FROM DIRECT EXPERIMENTS}

Direct laboratory experiments 
on neutrino magnetic moments
utilize solar, accelerator and reactor neutrinos
as sources, and are conducted under
controlled conditions.
These approaches are robust
and stay away from the ambiguities and model-dependence
in the astrophysical bounds.
The experiments require an understanding
of the neutrino energy spectrum as well as
its flavor/mass-eigenstate compositions at the detectors 
by independent means. 
They typically study neutrino-electron scatterings
$\nu_l + e \rightarrow \nu_x + e$.
The signature is
an excess of events
over those due to Standard Model (SM)
and other background processes, which
exhibit the characteristic
1/T spectral dependence.
Limits from negative searches 
are valid for both Dirac and Majorana
neutrinos and for all  final
states $\nu_x$, that is, for both
diagonal and transitional moments.  
However, comparisons and interpretations
among various experiments should take into
account the difference in the 
compositions between them {\it at} the
detectors.

\subsection{Solar Neutrinos}

Data from the solar
neutrino and KamLAND experiments
firmly established the validity of
the Standard Solar Model (SSM) predictions
of the solar neutrino flux, as well
as the LMA-matter oscillation solution being
the leading mechanism of neutrino flavor conversion
in the Sun. 
This can be used as the basis of magnetic
moment searches with solar neutrino data.

The Super-Kamiokande (SK) Collaboration performed
spectral distortion analysis of their
electron recoil spectral due to
$^8$B solar neutrino-electron scattering\cite{sk04}. 
The study was to look for 1/T excess over
an oscillation ``background'' at the 5-14~MeV energy range.
SK data alone allowed a large region
of ($\Delta m^2 , tan^2 \theta$) parameter space
and could only set limit of
$\munu ( ^8$B-$ \nu_{\odot} ) < 3.6 \times 10^{-10} \mub$
at 90\% Confidence Level (CL). 
Coupling with constraints from the other solar neutrino 
and KamLAND results,
the LMA region is uniquely selected as {\it the} solution,
such that a more stringent limit of
$\munu ( ^8$B-$ \nu_{\odot ; LMA} ) < 1.1 \times 10^{-10} \mub$
at 90\% CL  was derived.

The Borexino Collaboration performed analysis of
their Counting Test Facility data at the
$^7$Be solar neutrino relevant range:  
200$-$500~keV\cite{ctf03}.
Subtracting the known $^{14}$C $\beta$-spectrum and
{\it assuming} an additional linear background, 
a fit to look for an 1/T spectrum did not indicate
any excess and  a limit of 
$\munu ( ^7$Be-$ \nu_{\odot} ) < 5.5 \times 10^{-10} \mub$
at 90\% CL was derived, using SSM
$^7$Be $\nu_{\odot}$ flux.

An innovative insight is that neutrino
magnetic moments can induce photo-dissociation
in deuterium. 
The agreement between SNO neutral-current measurements
with SSM $\nu_{\odot}$-flux
predictions placed constraints to
the $\nu_e$-$d$ neutral-currents cross-sections
and thus to the magnetic moment effects:
$\munu ( ^8$B-$ \nu_{\odot} ) < 3.7 \times 10^{-9} \mub$
at 95\% CL\cite{snonud}.

\subsection{Accelerator Neutrinos}

Accelerators provide neutrinos with
known flavor compositions. 
The timing structures can be used for background 
subtraction. Compared to reactor neutrinos,
the lower flux as well as higher energy
limit the sensitivities. However, neutrinos
of all three flavors are produced at accelerators
such that this is the only laboratory avenue
for studying magnetic moments from $\nu_{\mu}$
and $\nu_{\tau}$.

The LSND experiment measured ``single electron''
events from a beam with known $\nu_e$,
$\nu_{\mu}$ and $\bar{\nu_{\mu}}$
fluxes and spectral compositions\cite{lsnd01}.
Taking the SM calculated values of
$\sigma ( \nu_{\mu}$-$ e )$
and 
$\sigma ( \bar{\nu_{\mu}}  $-$ e )$
which were confirmed by other experiments,
the value of $\sigma ( \nu_e  $-$ e )$
was derived. It agreed well with
SM predictions and provided a measurement
of $sin ^2 \theta_W = 0.248 \pm 0.051$.
Limits of
$\munu ( \nu_e ) < 1.1 \times 10^{-9} ~ \mub$
and
$\munu ( \nu_{\mu} ) < 6.8 \times 10^{-10} ~ \mub$
at 90\% CL
were derived from the absence of excess
of counts.

The DONUT experiment first observed 
explicit $\nu_{\tau}$ charged-current interactions\cite{pdg04}
showing that the $\nu_{\tau}$ flux at a beam dump configuration
is consistent with the expected level.
The experiment also looked for possible 
``single-electron'' events at cross-sections
much larger than SM expectations.
One event was observed while
the predicted background from
other known sources was 2.3\cite{donut01}.
This was used to convert into a
magnetic moment limit for  $\nu_{\tau}$:
$\munu ( \nu_{\tau} ) < 3.9 \times 10^{-7} ~ \mub$
at 90\% CL.

\subsection{Reactor Neutrinos}

Reactor neutrino experiments provide the
most sensitive laboratory searches for the
magnetic moments of $\nuebar$,
taking advantages of the
high $\nuebar$ flux, low $\enu$ and better experimental
control via the reactor ON/OFF comparison.
Neutrino-electron scatterings were first
observed in the pioneering experiment\cite{reines}
at Savannah River.
A revised analysis of the data by Ref~\cite{vogelengel}
with improved input parameters
gave a positive signature
consistent with the interpretation of a
finite $\munu$ at 
$\munu ( \nuebar ) \sim 2-4 \times 10^{-10} ~ \mub$.

\begin{figure}[ht]
\center
\includegraphics[height=5cm,width=6.5cm]{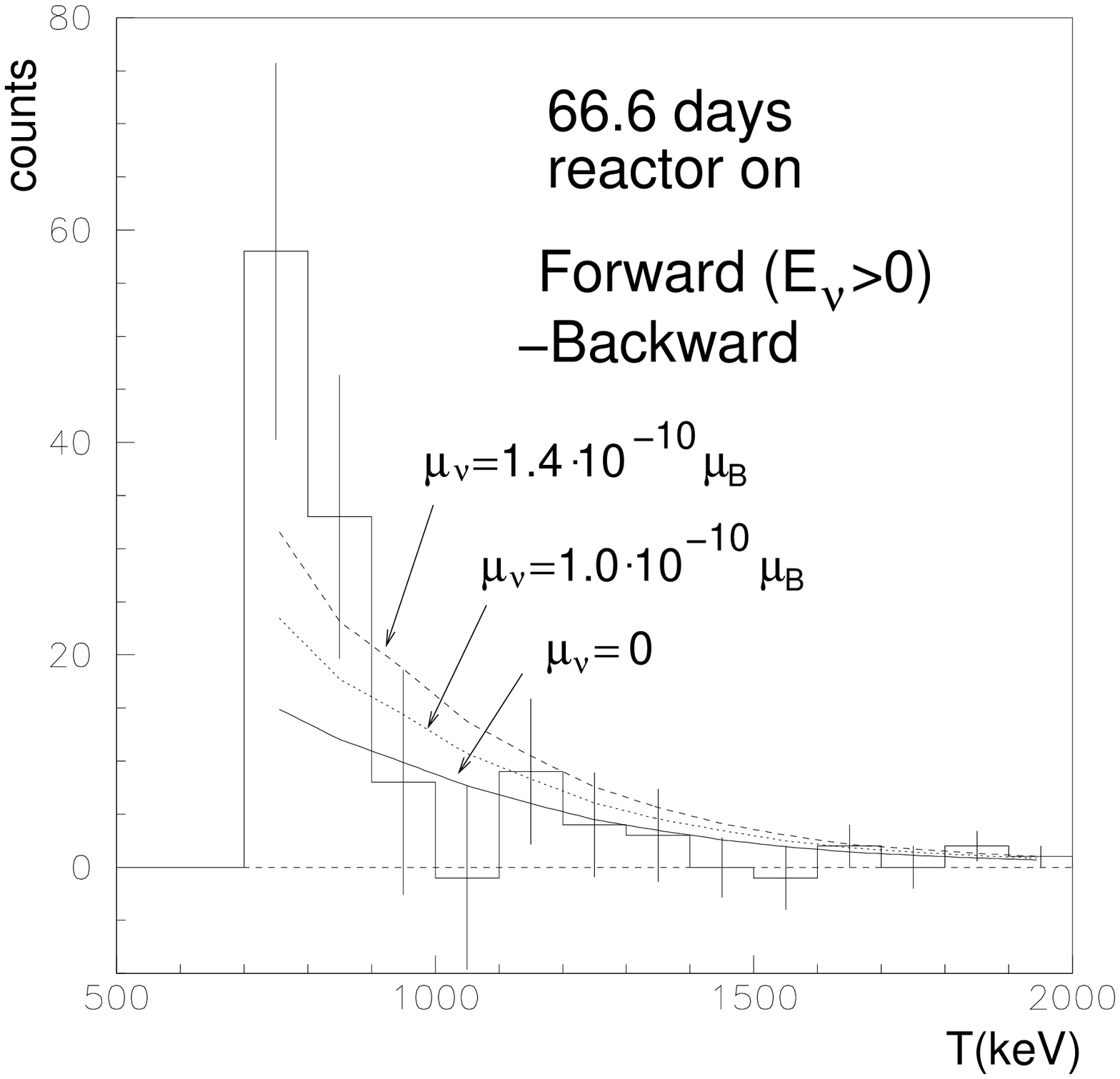}\\
\includegraphics[height=5cm,width=6.5cm]{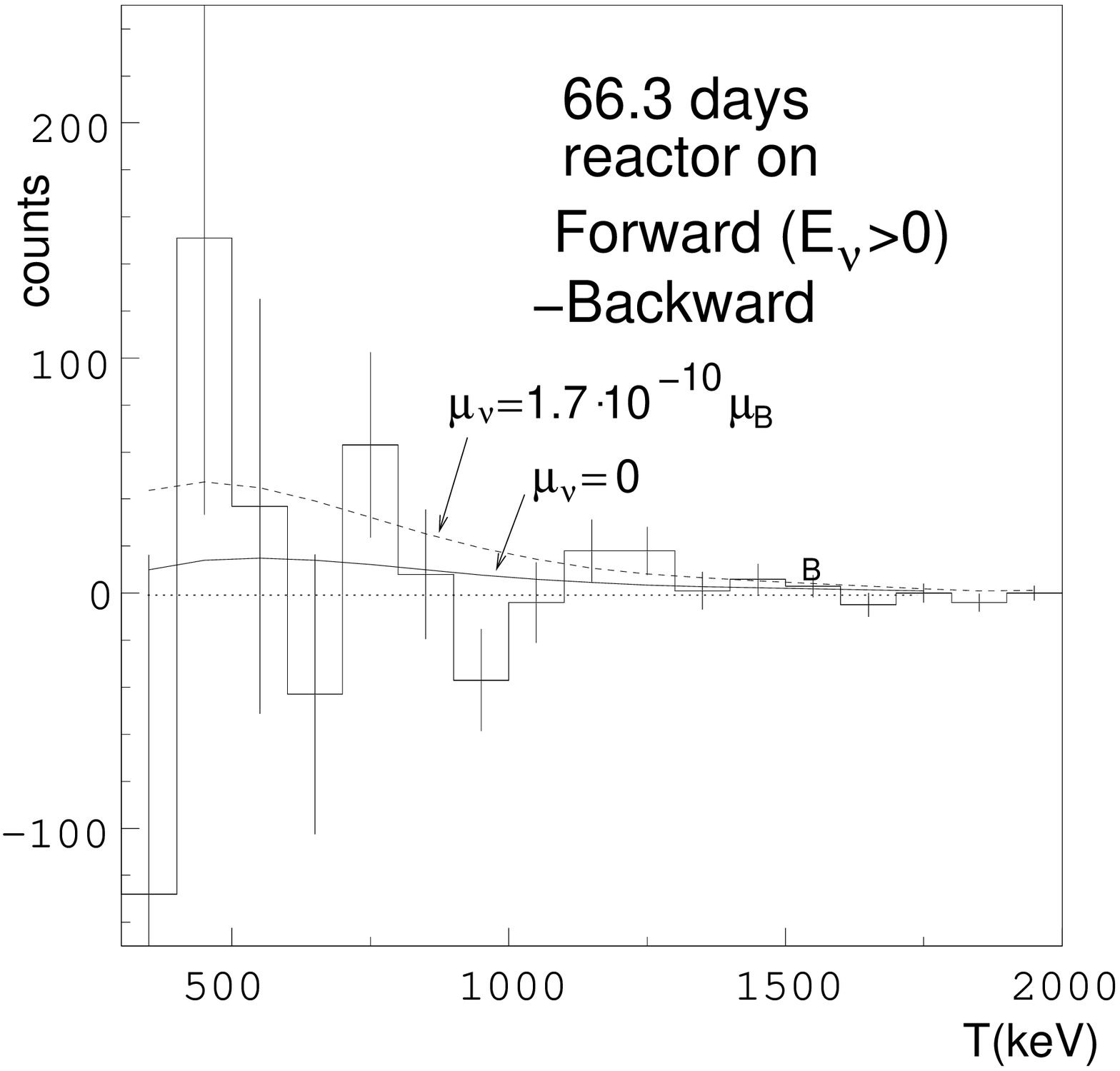}
\caption{
Residual spectra from MUNU for 
reactor-ON forward minus background events:
visual scan ($>$700 keV) and
automatic scan ($>$300 keV).
}
\label{munuresid}
\end{figure}

The MUNU experiment\cite{munu03}
at the Bugey reactor in France
deployed a Time Projection Chamber (TPC)
filled with CF$_4$ gas at 3~bar
having a mass of 11.4~kg, surrounded by
active liquid scintillator as anti-Compton
vetos. It gave
excellent single-electron event selection
and measured the scattering angle with respect
to the reactor core direction. 
Neutrino events are scattered ``forward'' 
such that a forward/backward comparison
was used to subtract background.
The residual spectra from 66.6~days of 
reactor ON data are depicted in Figure~\ref{munuresid}.
The residual counts
above T=900~keV
were consistent with SM expectations,
while
an excess of events
at 300~keV$<$T$<$900~keV
was observed
where the origins remain unknown.
Various limits were evaluated
depending on the analysis threshold:
$\munu ( \nuebar ) <  1.7/1.4/1.0 \times 10^{-10} ~ \mub$
at 90\% CL 
at 300, 700, 900~keV threshold, respectively.
The low energy ($<$2~MeV) reactor neutrino spectra
are not well-modeled\cite{lernu} such that the
possibility of yet-unaccounted-for
neutrino production channels at the MeV energy
range should be examined.

The TEXONO Collaboration adopted a
compact all-solid design with
an ultra-low-background high-purity
germanium (ULB-HPGe) detector with
a mass of 1.06~kg as target, surrounded
by anti-Compton detectors of NaI(Tl)
and CsI(Tl) crystal scintillators,
radon shields,
passive shielding materials, 
and cosmic-ray veto with
plastic scintillator panels.
The measurement was performed
at the Kuo-Sheng (KS) Power Plant in Taiwan.

The focus was on the T=10-100~keV range for
the enhanced signal rates and robustness
in the control of systematic uncertainties.
At this energy range, the $\nu$-e scattering
rates due to magnetic moments are much larger
than the SM rates at the $10^{-10} ~ \mub$
sensitivity level being explored, so
that uncertainties in the irreducible
SM background can be neglected\cite{lernu}. 
In addition, $T \ll E_{\nu}$ such that
the scattering rates due to
 $\munu$  depend on
the {\it total} neutrino flux ($\phi_{\nu}$)
rather than the poorly-known
details of the low-energy reactor neutrino
spectra. The total neutrino flux is
well-known and can be evaluated
accurately from reactor operation
data $-$
every fission is expected to produce
about 6 and 1.2 $\nuebar$'s due to 
$\beta$-decays of the fission daughters
and of $^{239}$U following neutron capture
on $^{238}$U, respectively.

\begin{figure}[ht]
\center
\includegraphics[width=6.5cm]{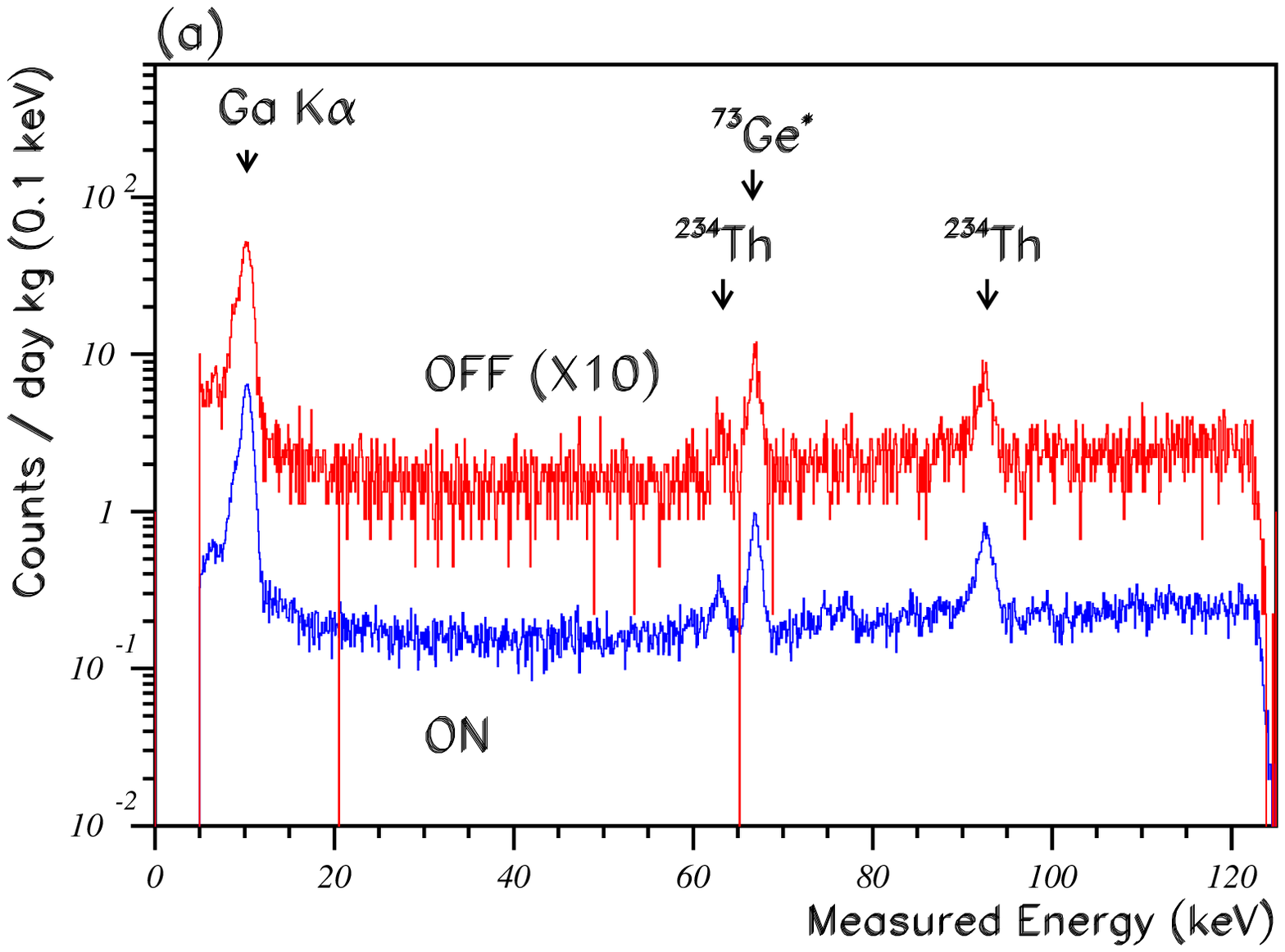}\\
\includegraphics[width=6.5cm]{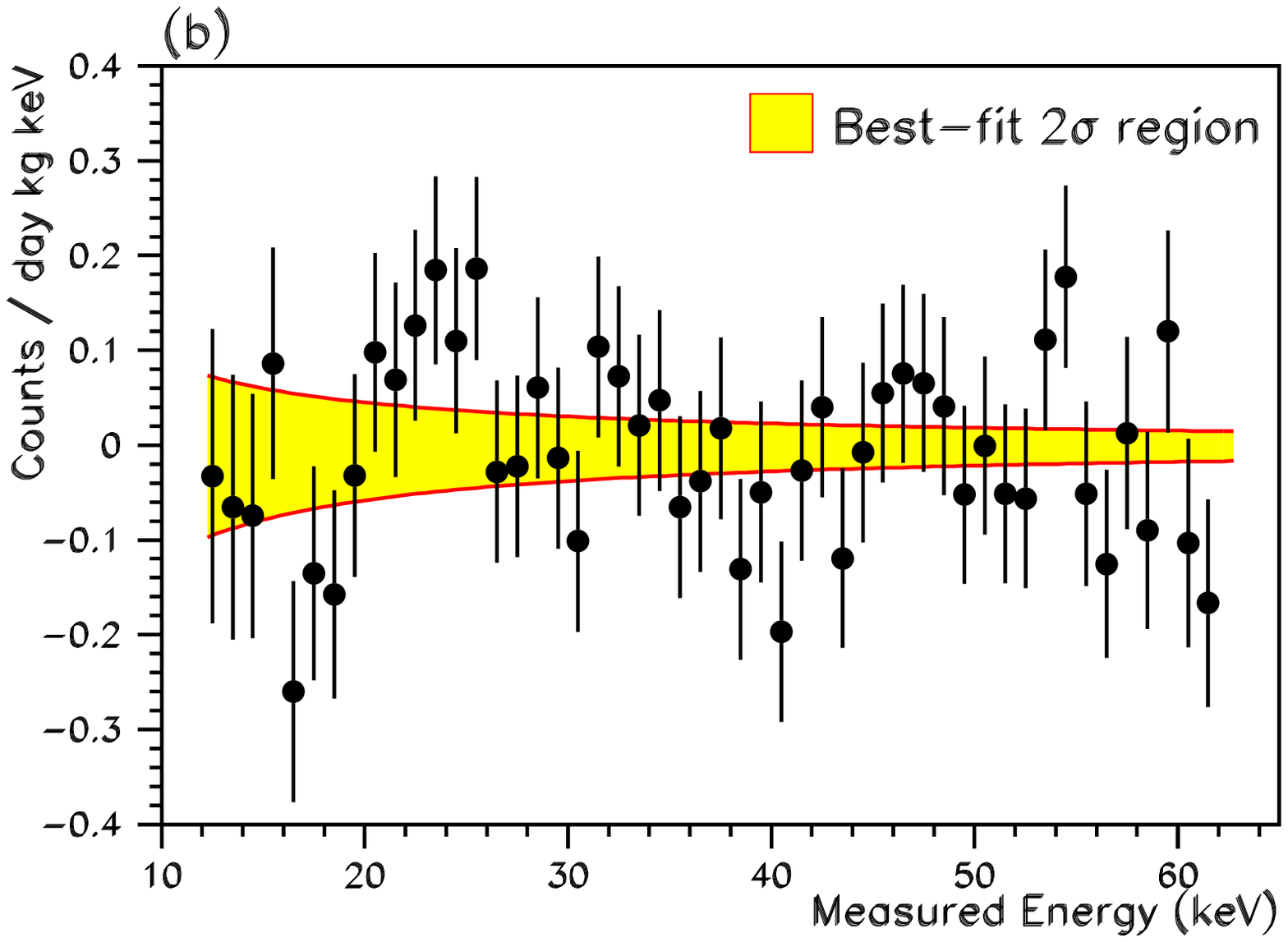}
\caption{
Data from TEXONO/KS on
(a) reactor ON/OFF energy spectra 
and (b) residual spectrum.
}
\label{texonoresid}
\end{figure}

Comparing 4712/1250 hours of reactor ON/OFF 
data, no excess of events was found 
and with an analysis threshold of 12~keV
just above the complications due to atomic effects,
a limit of 
$\munu ( \nuebar ) <  1.3 \times 10^{-10} ~ \mub$
at 90\% CL was derived.
The ON/OFF and residual spectra are displayed 
in Figure~\ref{texonoresid}.
Another notable result is that 
a background level
of $\sim 1 ~ kg^{-1} keV^{-1} day^{-1}$ 
at the 10-20~keV range was achieved $-$
a comparable range to those
from the underground Dark Matter experiments.

\begin{figure}[ht]
\center
\includegraphics[width=6.5cm]{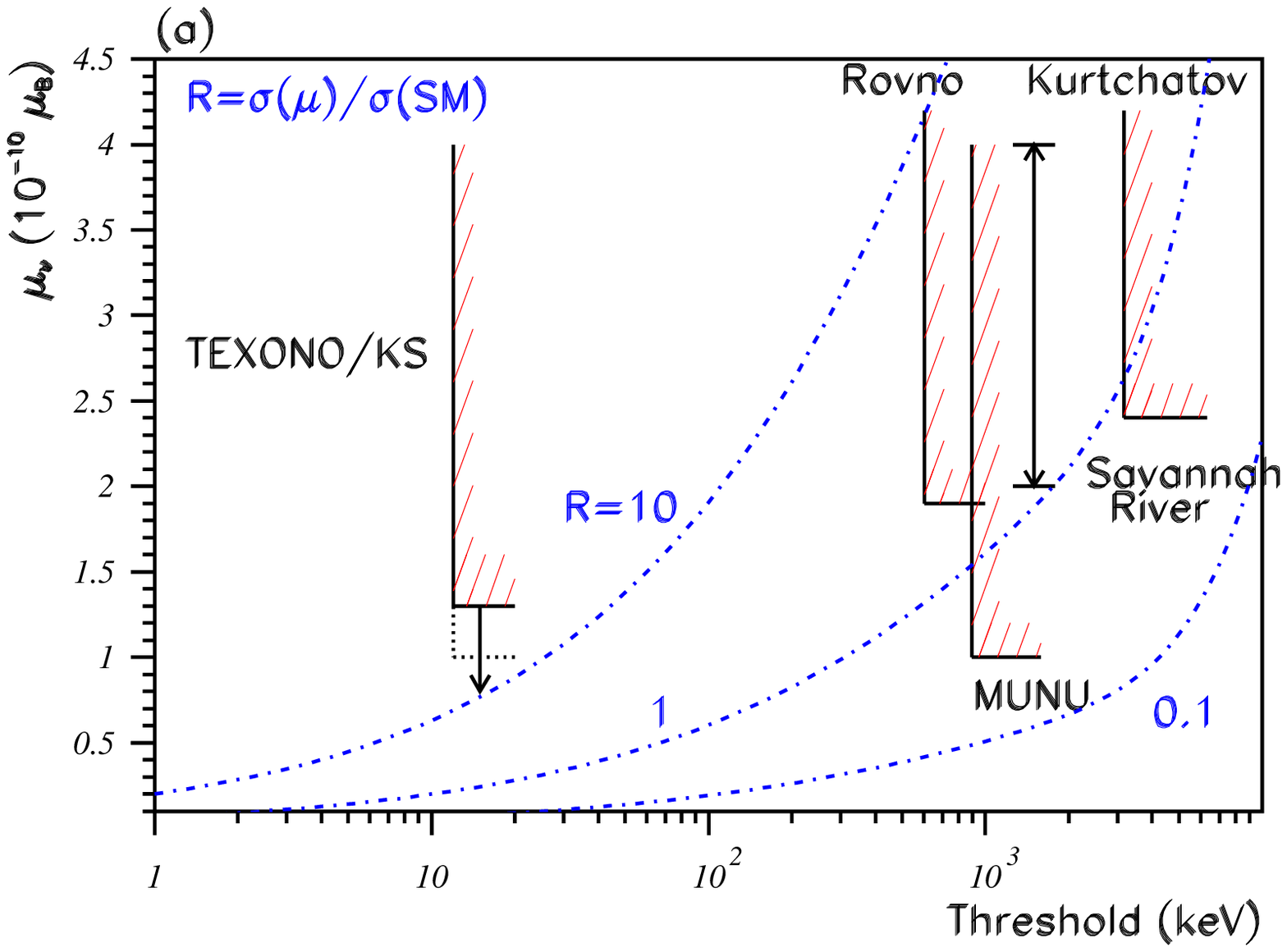}\\
\includegraphics[width=6.5cm]{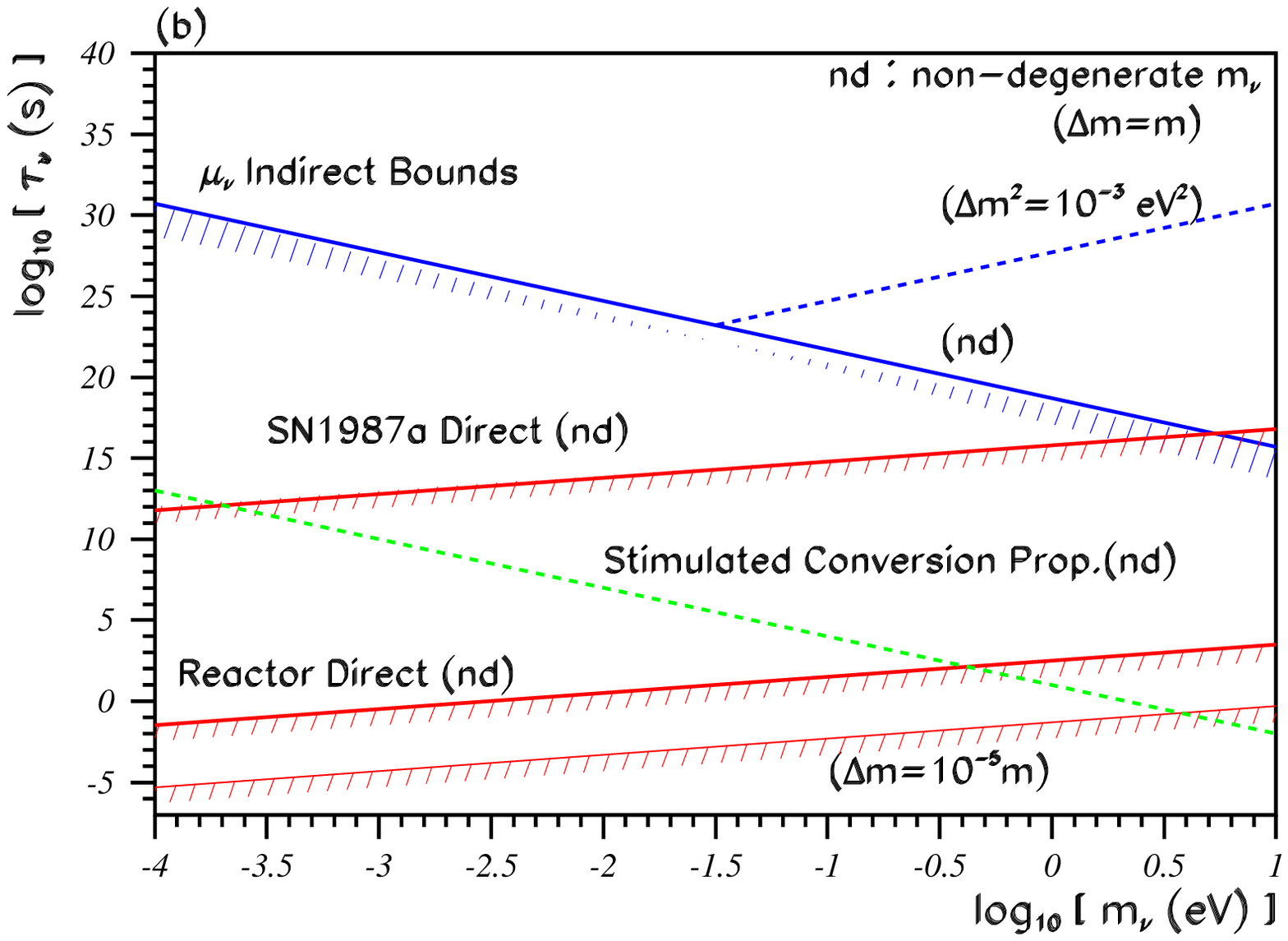}
\caption{
Summary of the results in
(a) the searches of neutrino
magnetic moments with reactor neutrinos,
and
(b) the bounds of
neutrino radiative decay lifetime.
}
\label{summaryplots}
\end{figure}

Depicted in Figure~\ref{summaryplots}a is the
summary of the results in $\rm{\munuebar}$ searches
versus the achieved threshold
in reactor experiments.
The dotted lines denote the
$\rm{R = \sigma ({\mu}) / \sigma (SM) }$ ratio at a
particular [T, $\rm{\munuebar}$].
The large R-values for the KS experiment
imply that its results
are robust against the uncertainties in the
SM cross-sections. In particular, in
the case where the excess events 
in Refs.~\cite{reines} and \cite{munu03} are due
to unaccounted sources of neutrinos, the
limits remain valid.
Indirect bounds on the neutrino radiative decay
lifetimes are inferred
and displayed in Figure~\ref{summaryplots}b
for the simplified scenario where a single channel
dominates the transition. It corresponds to
$\rm{
\tau_{\nu} m_{\nu} ^ 3 > 9.5 \times 10^{18} ~  eV ^3 s
}$ at 90\% CL in the non-degenerate case.
Superimposed are
the limits\cite{texono03}
from the previous direct searches
of excess
$\gamma$'s from reactor
and supernova SN1987a neutrinos, as well
as the sensitivities of proposed
simulated conversion experiments at accelerators.
It can be seen that
$\nu$-e scatterings give much more
stringent bounds than the direct approaches.

\subsection{Global Analysis}

A global analysis was performed\cite{global}
fitting simultaneously
the magnetic moment data from the
reactor and solar neutrino experiments, 
and the 
LMA oscillation parameters constrained by 
solar neutrino and KamLAND results.
Only Majorana neutrinos were considered such
that there were only transition moments.
A ``total'' magnetic moment vector 
$\Lambda = ( \mu_{23} , \mu_{31} , \mu_{12} )$
was defined, such that its
amplitude was given by
$ | \Lambda | ^ 2 = \frac{1}{2} Tr ( \mu ^{+} \mu ) $.
A global fit produced 90\% CL limits of
$  | \Lambda |  <  4.0 \times 10^{-10} ~ \mub$
from solar and KamLAND data,
and
$  | \Lambda |  <  1.8 \times 10^{-10} ~ \mub$
when reactor data were added.
The results indicate the role of
reactor experiments in constraining
the magnetic moment effects.

\section{FUTURE PROJECTS AND PROSPECTS}

The sensitivities for neutrino magnetic moments
in  direct search experiments
scale as
\begin{equation}
\munu \propto \frac{1}{\sqrt{N_{\nu}}} ~ [ \frac{B}{M ~  t} ]^{\frac{1}{4}}
\end{equation}
where $N_{\nu}$ is the signal events at some reference magnetic moments, 
$B,M,t$ are the background level, detector mass and measurement time,
respectively. It can be seen that the best strategy to
improve on the sensitivities is to increase on  $N_{\nu}$,
which is proportional to the neutrino flux $\phi_{\nu}$ and
is related to the detection threshold in recoil energy T.

The atomic energy level effects\cite{atomic} 
limit the potential enhancement of the
sensitivities by reducing T only. For example,
$N_{\nu}$ only increases by a factor of three 
with a lowering of detection threshold 
from 10~keV to 10~eV in Ge.
Therefore, big statistical boost in $\munu$ 
will most favorably be achieved by an 
enhancement in $\phi_{\nu}$ $-$ while 
keeping the systematics in control via
(a) lowering the detection threshold to retain
the ``$\munu \gg SM$'' event-rate requirements,
and (b) maintaining a low background level.
Since the minimal energy transfer to the atomic electrons
would be $\sim$100~eV, it follows from condition
(a) that such an approach 
of enhancing $\phi_{\nu}$ and reducing $T$
may only be applicable 
down to a $\munu$ sensitivity range of $10^{-13}~\mub$.

The GEMMA experiment\cite{gemma}
under preparation at the Kalininskaya
Nuclear Reactor in Russia is similar to
the TEXONO-KS approach, aiming at
an improvement to 
$\munu ( \nuebar ) \rightarrow  3 \times 10^{-11} ~ \mub$
by locating at a closer distance,
using a larger mass target and operating
at a lower threshold.
The MAMMONT project\cite{mamont},  
currently at the R\&D phase, 
has ambitious specifications of 
deploying a 40~MCi(4~kg) tritium
source with a flux of $6 \times 10^{14}~cm^{-2} s^{-1}$
on ultra-sensitive detectors with threshold down to
10~eV, either with cryogenic silicon detectors
or germanium with internal amplification.
The projected sensitivity is 
$\munu ( \nuebar ) \rightarrow  2.5 \times 10^{-12} ~ \mub$.

The TEXONO Collaboration continued data taking with
the ULE-HPGe at KS. Sensitivities to 
the $\sim 10^{-10} ~ \mub$
range can be expected. In parallel, a CsI(Tl) crystal
scintillator array\cite{texonocsi} with a total mass of 200~kg is also
collecting data. The strategy is to focus on the
high($>$3~MeV) recoil energy range to perform
a first measurement of SM neutrino-electron 
scattering at the MeV momentum transfer range.
A prototype ``Ultra-Low-Energy'' germanium 
detector with an active mass of 5~g is
being tested, with the goal of developing into
a 1~kg Ge-array detector for the first
experimental observation of neutrino-nucleus
coherent scattering. 
As by-product, such an experiment
will potentially probe 
$\munu ( \nuebar ) \rightarrow  2 \times 10^{-11} ~ \mub$.
An energy threshold
of 100~eV has been demonstrated\cite{texonocoh}
with the prototype while background studies 
at the sub-keV range are under way at KS.

Alternatives of neutrino
sources such as artificial radioactive sources\cite{nusources}
for NaI(Tl) and the Borexino detectors,
as well as accelerator-based 
$\beta$-sources for large TPCs\cite{betasources},
have been discussed, projecting a sensitivity
range of $\rm{\sim few \times 10^{-11}~\mub}$
in both cases.

\section{OUTLOOK}

The magnetic moments of the neutrino
parametrize how it couples to the
photons and are sensitive
to its masses and mixings, as well as
its Dirac or Majorana nature.
It is, therefore, a {\it conceptually}
rich subject with much neutrino
physics and astrophysics to be explored.
However, there are no indications of
any measurable/observable
positive signatures
in the current
and future rounds of experimental
efforts.
Improvement in
sensitivities will necessarily involve
new neutrino sources as well as
novel neutrino detection
techniques and channels. These advances
may find important potential applications
in other areas of neutrino
and underground physics experimentations.

\end{document}